\documentclass[pre,twocolumn,
showpacs,
amsmath,amssymb,
superscriptaddress
]{revtex4}

\newcommand{\be}{\begin{equation}}
\newcommand{\ee}{\end{equation}} 
\newcommand{\bea}{\begin{eqnarray}} 
\usepackage{graphicx}
\newcommand{\eea}{\end{eqnarray}}
\usepackage{epsfig}
\usepackage{bm} 

\begin{document}

\title{Conformal Invariance in (2+1)-Dimensional Stochastic Systems}
\author{L. Moriconi}
\affiliation{Instituto de F\'\i sica, Universidade Federal do Rio de Janeiro, \\
C.P. 68528, 21945-970, Rio de Janeiro, RJ, Brazil}
\author{M. Moriconi}
\affiliation{Instituto de F\'\i sica, Universidade Federal Fluminense, \\
Avenida Litor\^anea s/n, 24210-340, Niter\'oi, RJ, Brazil}

\begin{abstract}
Stochastic partial differential equations can be used to model second order thermodynamical phase transitions, as well as a number of critical out-of-equilibrium phenomena. In (2+1) dimensions, many of these systems are conjectured (and some are indeed proved) to be described by conformal field theories. We advance, in the framework of the Martin-Siggia-Rose field theoretical formalism of stochastic dynamics, a general solution of the translation Ward identities, which yields a putative conformal energy-momentum tensor. Even though the computation of energy-momentum correlators is obstructed, in principle, by dimensional reduction issues, these are bypassed by the addition of replicated fields to the original (2+1)-dimensional model. The method is illustrated with an application to the Kardar-Parisi-Zhang (KPZ) model of surface growth. The consistency of the approach is checked by means of a straightforward perturbative analysis of the KPZ ultraviolet region, leading, as expected, to its $c=1$ conformal fixed point.
\end{abstract}
\pacs{64.60.Ht, 64.60.De, 05.70.Jk}
\maketitle

Recent developments in two-dimensional criticality have extended the application of conformal field theory (CFT) methods \cite{bpz,difran_etal} to contexts far beyond the usual realm of second order phase transitions. Navier-Stokes turbulence \cite{bernard1,bernard2}, random deposition models \cite{saberi1, saberi2, saberi3, gier,alc-ritt}, spin glasses \cite{amoruso,bernard3} and even chaotic quantum wavefunctions \cite{keating,bogolmony} provide actual examples where conformality has been convincingly established. The theory of stochastic Lowner equations (SLE's) \cite{cardy,bauer,gruzberg} has been a major tool in the study of these problems \cite{comment1}. The essential strategy, which has strong geometrical appeal, is to identify level curves of order parameters or dynamical fields to the conformally invariant random paths described by SLE's. 
Nevertheless the clear success of numerical results, it is important to stress that they have been obtained in a frankly heuristic way, so that further elaborations are still in order.

The understanding of critical two-dimensional out-of-equilibrium systems under the light of conformal methods is still in its initial stages. 
The hope, supported by the above mentioned examples, is that conformal invariance will play a crucial role (as it does in thermodynamics) 
in extending the very concept of universality classes to the broad set of self-organized critical models (two-dimensional turbulence, 
for instance, turns out to be in the universality class of critical percolation \cite{bernard1}).

We put forward in this work an alternative, and essentially analytical, approach to the problem of conformal invariance in (2+1)-dimensional dynamical systems. As an interesting testing ground, we apply the method to the Kardar-Parisi-Zhang (KPZ) model of kinetic roughening \cite{kpz,saberi2}, a prototype of great interest in out-of-equilibrium statistical mechanics, due to its numerous connections to subjects like directed polymers in 
random media, Burgers hydrodynamics, flame fronts, domain wall dynamics, etc \cite{bs}.

Let $\phi =\phi(\vec x,t)$ be a general scalar field defined in $(2+1)$-dimensional spacetime. Assume that correlation functions of $\phi$ can be computed with the help of a path-integral measure $D \phi \exp(-S[\phi])$, not necessarily real. We are interested to know if spatial fluctuations of $\phi$ at a given time instant, say $t=0$, have an underlying CFT description. Taking $\phi^s (\vec x) \equiv \phi(\vec x,0)$, we introduce the reduced action $\bar S[\phi^s]$ -- the ``$\phi^s$-model" (to be contrasted to the original (2+1)-dimensional ``$\phi$-model") -- which, up to normalization factors, is obtained from
\be
\exp (- \bar S[\phi^s]) \equiv \int_* D \phi \exp(-S[\phi]) \ , \ \label{eq1}
\ee
where the symbol $\int_*$ stands for path-integration subject to the boundary condition $\phi(\vec x,0)=\phi^s(\vec x)$. 

One may wonder if it is possible to define an energy-momentum tensor for the two-dimensional $\phi^s$-model, as a way of probing its conformal structure from the computation of short-distance expansions \cite{bpz}. Since it is in general not known how to evaluate $\bar S[\phi^s]$, it would seem hopeless to concentrate any effort on the computation of the corresponding energy-momentum tensor. However, as we show below, the energy-momentum route to conformality is still viable in a pragmatical sense. Following Noether's theorem, let $T^\phi_{ij}$, with $i,j=1,2$, be the second order tensor related to the shift of the action $S[\phi]$ induced by time independent infinitesimal translations $\epsilon_j(\vec x)$ \cite{cardy2},
\be
S[\phi] \rightarrow S[\phi] -\frac{1}{2\pi} \int dx_1dx_2 \partial_i \epsilon_j T_{ij}^\phi \ . \ \label{eq3}
\ee
As $T^\phi_{ij}$ does not depend on the time variable, we call it the ``projected energy-momentum tensor".
It is not difficult to show that
\be
T^s_{ij} (\vec x)
\equiv \exp (\bar S[\phi^s]) \int_* D \phi \exp(-S[\phi]) T^\phi_{ij}(\vec x) \label{eq4}
\ee
does indeed solve the translation Ward identities of the $\phi^s$ model, viz.,
\bea
&&\langle \partial_j T_{ij}^s(\vec x) \phi^s(\vec x_1) \phi^s(\vec x_2)...\phi^s(\vec x_N) \rangle_{\phi^s} 
\nonumber \\
&=& 2 \pi \sum_{p=1}^N \delta^2(\vec x -\vec x_p) \partial^{(p)}_i \langle \phi^s(\vec x_1) \phi^s(\vec x_2)...\phi^s(\vec x_N) \rangle_{\phi^s} 
\ . \  \nonumber \\ \label{eq5}
\eea
We will, thus, pursue the idea that Eq. (\ref{eq4}) can be used to define the renormalized energy-momentum tensor associated to the $\phi^s$-model \cite{comment2}. It is not possible to compute $T^s_{ij}$ in general, since it depends, according to (\ref{eq1}) and (\ref{eq4}), on exact path-integrations. Observe, however, that (\ref{eq4}) leads to
\bea
&&\int D \phi^s \exp (- 2 \bar S[\phi^s]) T^s_{ij} (x_1,x_2) T^s_{lm} (x_1',x_2') \nonumber \\
&&= \int D \phi_1 D \phi_2 \delta [\phi^s_1-\phi^s_2]  \exp ( -S[\phi_1]-S[\phi_2]) \nonumber \\
&&\times T^{\phi_1}_{ij}(x_1,x_2) T^{\phi_2}_{lm}(x_1',x_2') \ . \ \label{eq6}
\eea
The fields $\phi_1$ and $\phi_2$ in (\ref{eq6}) are replicas of $\phi$, which are path-integrated under the constraint $\phi_1(\vec x,0)=\phi_2(\vec x,0)$. Eq. (\ref{eq6}) would be exactly what one needs for the computation of the energy-momentum correlator, if the exponential factor in its LHS were given by $\exp(-\bar S[\phi^s])$. 

It is clear that the critical surface is modified due to the replacement of $\exp(-\bar S[\phi^s])$ by $\exp(-2 \bar S[\phi^s])$ in the path integration measure. However, since the dimensionality, symmetries and the form of interactions are strictly the same in both of these models, it is likely that the critical surface can be restored to its original shape through coupling constant redefinitions. In the case of self-organized critical dynamics (as in the KPZ model) it is natural to conjecture stability of the critical behavior against such critical surface deformations. 

As far as we are interested in criticality, Eq. (\ref{eq6}) is worth of attention due to the computability of its RHS, which avoids the use of the exact dimensionally reduced action $\bar S[\phi^s]$. Taking $w=x_1+ix_2$ and $T_\phi \equiv T_{11}-T_{22}+2iT_{12}$, the holomorphic component of the projected energy-momentum tensor derived in the $\phi$-model, conformality of the $\phi^s$-model will be indicated by 
\be
\langle T_{\phi_1}(w) T_{\phi_2}(w') \rangle = \frac{c}{8} \frac{1}{(w-w')^4} \ , \ \label{eq7}
\ee
where $c$ is the central charge \cite{bpz} and the above expectation value is computed from the path integration over fields $\phi_1$ and $\phi_2$ in (\ref{eq6}). The $1/8$ factor in (\ref{eq7}) is due to the fact that the energy-momentum tensor associated to the model with action $2 \bar S[\phi^s]$ is $2 T^s_{ij}$.

In order to understand how the method works in practice, we focus our attention, now, on its application to a specific problem: 
the (2+1)-dimensional KPZ model,
\be
\partial_t h = \nu \partial^2 h + \frac{\lambda}{2} (\partial_\alpha h)^2 + \sqrt{2D} \eta \ , \ \label{eq8}
\ee
which describes the evolution of the height $h$ of randomly deposited atoms on a planar substrate. The system is characterized by the surface tension $\nu$, the non-linear parameter $\lambda$ of lateral growth, and the strength of random deposition $D$. The gaussian random field $\eta$ is taken to have zero mean and two-point correlation function
$\langle \eta(\vec x,t) \eta(\vec x',t) \rangle = \delta^2(\vec x - \vec x') \delta(t-t') $.

The Cole-Hopf transformation, 
\be
\phi = \exp(\lambda h / 2 \nu ) \ , \ \label{eq9}
\ee
maps Eq. (\ref{eq8}) into
\be
\partial_t \phi = \partial^2 \phi + \sqrt{2g} \eta \phi \ , \ \label{eq10}
\ee
where $g = 2 \lambda^2 D /\nu^3$. We can write down, using the response functional formalism \cite{cardy2,hansen,dominicis}, the generating functional of correlation functions for the replicas $\phi_1$ and $\phi_2$ of model (\ref{eq10}),
\bea
&&Z[\{j_p, \hat j_p \}] = \int D \chi  D \hat \phi_p D \phi_p \nonumber \\
&&\times \exp \left  \{ - S + i\int dt d^2 \vec x (j_p \phi_p + \hat j_p \hat \phi_p)  \right \}
\ , \ \label{eq11}
\eea
where $S=S_\phi+S_\chi$ with
\bea
S_\phi &=& \int d^2 \vec x dt [i\hat \phi_p (\partial_t \phi_p - \partial^2 \phi_p) + g \hat \phi_p^2 \phi_p^2] \ , \ \label{eq12} \\
S_\chi &=& i \int d^2 \vec x \chi(\vec x) [\phi_2(\vec x,0)-\phi_1(\vec x,0)] \ . \ \label{eq13}
\eea
Above, $\chi(\vec x)$ is the Lagrange multiplier field which ensures,  as in (\ref{eq6}), the constraint
$\phi_1(\vec x,0) = \phi_2(\vec x,0)$. The system is confined, in principle, to a box of dimensions $L \times L$, and time integration 
is restricted to the finite interval $-\Delta \leq t \leq 0$, with eventually $L, \Delta \rightarrow \infty$ (the order of taking
limits can be important here).

Assuming that $h=0$ at initial time ($t=-\Delta$), the KPZ description is effectively a perturbation of the Edwards-Wilkinson (EW) linear model at small spacetime scales, where the surface height fluctuates around $\bar h =0$ \cite{fda}. We can perform, accordingly, the substitution $\phi \rightarrow 1+\phi$ in the action $S_\phi$, to get
\be
\partial_t \phi = \partial^2 \phi + \sqrt{g} \eta+\sqrt{g} \eta \phi \ . \ \label{eq14}
\ee
Perturbative renormalization group flows computed from either the stochastic Eqs. (\ref{eq8}) or (\ref{eq14}) agree perfectly well \cite{kpz,fns,bs,wiese,canet}, and show that $g=0$ is an ultraviolet fixed point in $d=2$. Any finite $g$ flows in the infrared to the (still barely understood) KPZ strong coupling regime \cite{canet}.

As usual, the generating functional can be split into quadratic and interacting parts. 
Considering the small scale model (\ref{eq14}), exact integrations over $\hat \phi$ and $\phi$ 
yield the non-perturbed functional
\bea
&&Z_0[\{J_q^p  \}] = \int D \chi
\exp \left \{ \frac{i}{4} \int dt d^2\vec x \int dt' d^2 \vec x' \right. \nonumber \\
&\times&
\left. \vphantom{\int} J_q^p(\vec x,t) J_r^p(\vec x',t')A_{q r}(\vec x, \vec x', t,t') \right \}
\ , \ \label{eq15}
\eea
where $J_q^p (\vec x,t)$ is the $q^{th}$ component of the doublet
\be
\mathbb{J}^p (\vec x,t) = 
\left[
\begin{array}{c}
\hat j_p (\vec x,t) \\
j_p(\vec x,t)-(-1)^p \chi (\vec x) \delta(t) \\
\end{array}
\right] \label{eq16}
\ee
and $A_{q r}(\vec x, \vec x', t,t')$ 
denotes the matrix elements of the operator
\be
\mathbb{A} = 
\frac{2}{\partial^2_t-\partial^4}
\left[
\begin{array}{cc}
0&- \partial_t +\partial^2 \\
\partial_t +\partial^2&-2ig \\
\end{array}
\right] \ , \  \label{eq17}
\ee
that is,
\bea
&&A_{12}(\vec x, \vec x', t, t') = A_{21} (\vec x, \vec x', t', t) = \nonumber \\
&&=\frac{1}{2 \pi} \frac{\Theta(t'-t)}{|t'-t|}
\exp \left [ - \frac{(\vec x - \vec x')^2}{4 |t'-t|} \right ] \nonumber \ , \ \\
&&A_{22}(\vec x, \vec x', t, t') = -\frac{ig}{\pi} \ln|\vec x -\vec x'| \nonumber \\
&& -\frac{ig}{ 2 \pi} \int_0^{t-t'} 
\frac{d \xi}{|\xi|}
\exp \left [ - \frac{(\vec x - \vec x')^2}{4 |\xi|}  \right ] \ . \ \nonumber \\
\label{eq18}
\eea
The complete generating functional is written, then, as
\bea
&&Z[\{J_q^p  \}] = \nonumber \\
&&= \exp \left \{ \int dt d^2 \vec x \left [ 2 g 
\frac{\delta^3}{\delta \hat j_p^2 \delta j_p} 
- g \frac{\delta^4}{\delta \hat j_p^2 \delta j_p^2} \right ] \right \} 
Z_0[\{J_q^p  \}] \ . \ \nonumber \\
\label{eq19}
\eea
Applying time-independent spatial translations to the
$\phi$-model's action, the holomorphic projected energy-momentum tensor 
is readily computed as
\be
T(w) = 4 \pi i \int dt \partial_w \hat \phi \partial_w \phi \ . \ \label{eq20}
\ee
The time variable is extended here to $-\infty < t < \infty$. 
Note that due to the assumption of statistical stationarity one does not have to worry about 
boundary conditions at $t \rightarrow \pm \infty$.

Since the renormalized $g$ vanishes at small scales, the exponential in (\ref{eq19}) plays no role, and we expect the energy-momentum correlator to be given, in the ultraviolet region, by the free (quadratic) approximation:
\bea
{\hspace{-0.5cm}} &&\langle T_1(w) T_2(w') \rangle = -(4 \pi)^2 \int dt \int dt' \nonumber \\
&&{\hspace{-0.5cm}}\times [ \langle \partial_w \phi_1(w,t) \partial_{w'} \phi_2(w',t') \rangle_0 
\langle \partial_w \hat \phi_1(w,t) \partial_{w'} \hat \phi_2(w',t') \rangle_0 \nonumber \\
&&{\hspace{-0.5cm}}+ \langle \partial_w \hat \phi_1(w,t) \partial_{w'} \phi_2(w',t') \rangle_0 
\langle \partial_w \phi_1(w,t) \partial_{w'} \hat \phi_2(w',t') \rangle_0 ] \ . \ \nonumber \\
\label{eq21}
\eea
The above two-point correlation functions are computed from the functional derivatives of $Z_0[\{J^p_q\}]$ at vanishing currents. We obtain
\bea
&&\langle \phi_1(w,t) \phi_2(w',t') \rangle = \nonumber \\
&&= \frac{1}{8 \pi^2} \int d^2 \vec k \frac{1}{k^2}
\exp[-k^2(|t|+|t'|)+i\vec k \cdot (\vec x - \vec x')] \ , \ \nonumber \\
&&\langle \hat \phi_1(w,t) \hat \phi_2(w',t') \rangle = - \Theta(-t) \Theta(-t') \times \nonumber \\
&& \times \frac{1}{8 \pi^2} \int d^2 \vec k k^2
\exp[-k^2(|t|+|t'|) +i\vec k \cdot (\vec x - \vec x')] \ , \ \nonumber \\
&&\langle \hat \phi_1(w,t) \phi_2(w',t') \rangle =  \nonumber \\
&& -\frac{ i\Theta(-t)}{8 \pi^2} \int d^2 \vec k
\exp[-k^2(|t|+|t'|) +i\vec k \cdot (\vec x - \vec x')] \ . \ \nonumber \\
\eea
Note that $\langle \hat \phi_1(w,t) \phi_2(w',t') \rangle = \langle \hat \phi_2(w,t) \phi_1(w',t') \rangle$. 
We are led, in this way, after a straightforward but lengthy computation, to
\bea
\langle T_1(w) T_2(w') \rangle &=& \frac{1}{32} \left [ \int_0^\infty dr r J_2(r) \right ]^2 \frac{1}{(w-w')^4} \nonumber \\
&=& \frac{1}{8} \frac{1}{(w-w')^4} \ , \ \label{eq22}
\eea
where $J_2(r)$ is the second order Bessel function of the first kind. It is remarkable that (\ref{eq22}) does not depend on
the coupling constant $g$ at all (for finite $g$). Comparing (\ref{eq7}) and (\ref{eq22}), we find out that in the EW regime of the KPZ model, surface growing is conformal, with unit central charge, in agreement with the numerical SLE results \cite{saberi2}. 

At larger spacetime scales, the EW regime breaks down, as indicated by anomalous exponents in the Family-Vicsek (FV) finite size scaling relation for the roughness, $W \equiv \sqrt{ \langle (h-\bar h)^2 \rangle } = L^\alpha F(t/L^z)$ \cite{bs,fv}. In two dimensions, EW exponents are exactly $\alpha = 0$ and $z=2$ \cite{bs}, while for the KPZ model, $\alpha \simeq 0.38$ and $z \simeq 1.62$ \cite{cola-moore, fda2}. 

Unfortunately, we are unable to access, by means of perturbation theory, the strong coupling (large scale) regime of the surface growing model (\ref{eq10}). The numerical analysis of Ref. \cite{saberi2} points out that the KPZ universality class is conformal as well, with vanishing central charge. It is a challenging problem to derive it along the analytical formalism presented here. It is worth mentioning that there are at least two non-perturbative approaches that could be useful, in principle, to address evaluations of the energy momentum correlator in the KPZ strong coupling regime: one is the ``exact renormalization group" technique \cite{canet}; the other is the $1/N$ expansion developed in \cite{doherty_etal}.

To summarize, we have discussed the issue of conformal invariance in (2+1)-dimensional stochastic systems within a systematic field-theoretical framework. A dimensionally reduced energy-momentum tensor is defined in two dimensions from the original dynamical model, as a solution of translation Ward identities. Our main working hypothesis is that critical surfaces are not spoiled by the introduction of replicated auxiliary fields, which avoid computational complications associated with dimensional reduction. The approach opens the way for analytical studies of conformality in a broad class of out-of-equilibrium models. A performance test has been carried out for the short-distance (weak-coupling) regime of the KPZ model of surface growing, where the method is noted to work with perfection.

\acknowledgements
This work has been partially supported by CNPq and FAPERJ. We thank F.D.A. Aar\~ao Reis for
valuable comments.


\begin{references}
\bibitem{comment1} We note that Refs. \cite{gier} and \cite{alc-ritt} establish conformal invariance 
in stochastic (1+1)-dimensional models in a purely algebraic way, without the use of SLE methods.
\bibitem{comment2} Eq. (\ref{eq4}) can be easily proved in relativistic models, taking functional derivatives of 
(\ref{eq1}) with respect to the metric components.
\end{references}
\end{document}